\documentclass[12pt]{iopart}

\usepackage{graphics,epsfig,subfigure}

\begin{document}

\title[]{A Note on Temperature and Energy of 4-dimensional Black Holes from Entropic Force}

\author{Yu-Xiao Liu,
        Yong-Qiang Wang\footnote{Corresponding author.},
        Shao-Wen Wei}
\address{
    Institute of Theoretical Physics,
    Lanzhou University,
    Lanzhou 730000, P. R. China}
\ead{liuyx@lzu.edu.cn, yqwang@lzu.edu.cn and weishaow06@lzu.cn}

\begin{abstract}
We investigate the temperature and energy on
holographic screens for 4-dimensional black holes with the
entropic force idea proposed by Verlinde. We find that the
``Unruh-Verlinde temperature'' is equal to the Hawking
temperature on the horizon and can be considered as a
generalized Hawking temperature on the holographic screen
outside the horizons. The energy on the holographic screen
is not the black hole mass $M$ but the reduced mass $M_0$,
which is related to the black hole parameters. With the
replacement of the black hole mass $M$ by the reduced mass
$M_0$, the entropic force can be written as
$F=\frac{GmM_0}{r^2}$, which could be tested by
experiments.

\end{abstract}

\maketitle

\section{Introduction}\label{Sec1}

Recently, Verlinde \cite{Verlinde} presented a remarkable new idea that gravity
can be explained as an entropic force caused by the information changes when a
material body moves away from the holographic screen. This idea implies that
gravity is not fundamental. With the holographic principle and the
equipartition law of energy, Verlinde showed that Newton's law of gravitation
can arise naturally and unavoidably in a theory in which space is emergent
through a holographic scenario, and a relativistic generalization leads to the
Einstein equations. In fact, the similar idea can be traced back to Sakharov's
work \cite{Sakharov}. On the other side, using the equipartition law of energy
for the horizon degrees of freedom together with the thermodynamic relation
$S=E/(2T)$, Padmanabhan also obtained the Newton's law of gravity
\cite{Padmanabhan1,Padmanabhan2}.

Subsequently, with the idea of entropic force, some applications have
been carried out. The Friemdann equations and the modified Friedmann
equations for Friedmann-Robertson-Walker universe in Einstein gravity
\cite{Y.G.Gong1,R.G.Cai}, $f(R)$ gravity \cite{Y.G.Gong2}, deformed
Horava-Lifshitz gravity \cite{WeiLiu1}, and braneworld scenario
\cite{Y.Ling} were derived with the help of holographic principle and
the equipartition rule of energy. The Newtonian gravity in loop
quantum gravity was derived in Ref. \cite{Smolin}. In Ref.
\cite{T.Wang} the coulomb force was regarded as an entropic force. In
Ref. \cite{M.Li}, it was shown that the holographic dark energy can
be derived from the entropic force formula. It was pointed out in
\cite{Makela} that Verlinde's entropic force is actually the
consequence of a specific microscopic model of spacetime. The similar
ideas were also applied to the construction of holographic actions
from black hole entropy \cite{Caravellila}. While Ref.
\cite{J.-W.Lee} showed that gravity has a quantum informational
origin.

On the other hand, a modified entropic force in the Debye's model was gave in
\cite{Gao}. In Ref. \cite{L.Zhao}, some of the problems of Verlinde's proposal
on the thermodynamical origin of the principle of relativity was presented, and
the thermodynamic origin of the principle of relativity was explained by hidden
symmetries of thermodynamics. Other applications can be seen in
\cite{Piazza,Culetu,Wang2,WuXN1002.1275}.

According to Verlinde's idea, the holographic screens locate at equipotential
surfaces, where the potential is defined by a timelike Killing vector. Then the
local temperature on a screen can be defined by the acceleration of a particle
that is located very close to the screen. The energy on the screen is
calculated by the holographic principle and the equipartition rule of energy
$E=\frac{1}{2}{\int}TdN$ with $dN$ the bit density of information on the
screen. In this paper, we apply these formulas to investigate the temperature
and energy on holographic screens for 4-dimensional static spherically
symmetric and the Kerr black hole, and look for experiment methods for testing
the entropic force.

The paper is organized as follows. In Sec. \ref{Sec2}, we review Verlinde's
idea about the temperature and the energy from an entropic force in general
relativity. Then, with the idea of entropic force together with the
equipartition law of energy, we calculate the temperature and the energy for
4-dimensional static spherically symmetric black holes and the Kerr black hole
in Sec. \ref{Sec3} and Sec. \ref{Sec4}, respectively. Finally, we give a brief
discussion and present that the entropic force could be tested by experiments.

\section{ The temperature and the energy from an entropic force}
\label{Sec2}

We first review the idea of Verlinde about the temperature and the
energy from an entropic force in general relativity. Consider a
static background with a global time like Killing vector $\xi^\mu$.
One can relate the choice of this Killing vector field with the
temperature and the energy.

In order to define the temperature, we first need to introduce the
potential $\phi$ via the timelike Killing vector $\xi^\mu$:
\begin{eqnarray}
  \phi ={1\over 2} \log(-\xi^\mu\xi_\mu),\label{phi}
\end{eqnarray}
where $\xi_\mu$ satisfies the Killing equation
\begin{eqnarray}
\nabla_\mu\xi_\nu+\nabla_\nu\xi_\mu=0.
\end{eqnarray}
The redshift factor is denoted by $e^\phi$ and it relates the local
time coordinate to that at a reference point with $\phi=0$, which we
will take to be at infinity. The potential is used to define a
foliation of space, and the holographic screens are put at surfaces
of constant redshift. So the entire screen has the same time
coordinate.

The four velocity $u^\mu$ and the acceleration
$a^\mu\!\equiv\!u^\nu\nabla_\nu u^\mu$ of a particle that is located
very close to the screen can be expressed in terms of the Killing
vector $\xi^\mu$ as
\begin{eqnarray}
 u^\mu &=& e^{-\phi}\xi^\mu, \\
 a^\mu &=& e^{-2\phi}\xi^\nu\nabla_\nu\xi^\mu
     =-\nabla^{\mu}\phi. \label{amu}
\end{eqnarray}
Note that because the acceleration is the gradient of the potential, it is
perpendicular to screen $\cal S$. So we can turn it into a scalar quantity by
contracting it with a unit outward pointing vector $n^\mu$ normal to the screen
$\cal S$ and to $\xi^\mu$. With the normal vector
$n^\mu=\nabla^\mu\phi/\sqrt{\nabla^\nu\phi\nabla_\nu\phi}$, the local
temperature $T$ on the screen is defined by
\begin{eqnarray}
  T=-{\hbar \over 2\pi} e^{\phi} n^\mu a_\mu
   ={\hbar \over 2\pi} e^{\phi} n^\mu\nabla_\mu\phi
   ={\hbar \over 2\pi} e^{\phi} \sqrt{\nabla^\mu\phi\nabla_\mu\phi},
   \label{T}
\end{eqnarray}
where a redshift factor $e^\phi$ is inserted because the
temperature $T$ is measured with respect to the reference point at
infinity. We will call the temperature defined in (\ref{T}) as
Unruh-Verlinde temperature.

Assuming that the change of entropy at the screen is $2\pi$ for a
displacement by one Compton wavelength normal to the screen, one has
 \begin{eqnarray}
 \nabla_\mu S = -2\pi {m\over \hbar} n_\mu.
 \end{eqnarray}
Now the entropic force, which is required to keep a particle with mass $m$ at
fixed position near the screen, is turned out to be
 \begin{eqnarray}
 \label{entropicforce}
 F_\mu =T\nabla_\mu S = -m e^\phi\nabla_\mu\phi,\label{Fmu}
 \end{eqnarray}
where $-\nabla_\mu\phi$ is the relativistic analogue of Newton's acceleration,
and the additional factor $e^\phi$ is due to the redshift, which is a
consequence of the entropy gradient.

We now consider a holographic screen on a closed surface of
constant redshift $\phi$. The number of bit $N$ of the
screen is assumed to proportional to the area of the screen
and is given by \cite{Padmanabhan2010a}
\begin{eqnarray}
 N= {A\over \hbar}.
\end{eqnarray}
Then, by assuming that each bit on the holographic screen contributes
an energy  $T/2$ to the system, and by using the equipartition law of
energy, we get
\begin{eqnarray} \label{Energy_a}
 E={1\over 2} \int_{\cal S}  T dN
\end{eqnarray}
with $dN$ the bit density on the screen. Inserting the expressions of
$T$ and $N$ into (\ref{Energy_a}) results in
\begin{eqnarray} \label{Energy}
 E ={1\over 4\pi}\int_{\cal S} e^{\phi} \nabla \phi ~ dA.\label{E}
\end{eqnarray}

In the following, we will calculate the Unruh-Verlinde temperature
and the energy on holographic screens for 4-dimensional black
holes with static spherically symmetric metric and stationary
axisymmetric metric, respectively.

\section{Unruh-Verlinde Temperature and energy for static spherically symmetric black holes}
 \label{Sec3}

We first consider the case of 4-dimensional general static
spherically symmetric black holes. In the general case, the metric
can be taken as the form
\begin{equation}
 ds^2 = -f(r) dt^2 + \frac{dr^2}{f(r)} + r^2 d\Omega_2^2, \label{general1}
\end{equation}
where $d\Omega_2^2 =d\theta^2+\sin^{2}{\theta}d\varphi^2$ is the metric on a
unit 2-sphere. We assume
\begin{equation}
 \lim_{r \rightarrow \infty} f(r) = 1 \label{assume_f(r)}
\end{equation}
for ensuring that the metric is asymptotically flat. The event
horizon radius $r_{EH}$ is usually determined by the largest
solution of $f(r)=0$.

By using the Killing equation
\begin{equation}
 \partial_\mu\xi_\nu +\partial_\nu\xi_\mu
 - 2 \Gamma^{\lambda}_{\mu\nu} \xi_{\lambda}=0,\label{KillingEq}
\end{equation}
and the static spherically symmetric properties of the metric
(\ref{general1})
\begin{equation}
 \partial_t \xi_\mu = \partial_\varphi \xi_\mu=0,
\end{equation}
as well as the condition $\xi_\mu\xi^\mu=-1$ at infinity, the
timelike Killing vector of the general black hole is solved as
\begin{eqnarray}
 \xi_\mu=\left(-f(r),0,0,0\right),
\end{eqnarray}
which is zero at the event horizon.

According to (\ref{phi}), (\ref{amu}) and (\ref{T}), the
potential, the acceleration and the Unruh-Verlinde temperature on
the holographic screen put at the spherical surface with radius
$r$ are calculated as \footnote{We are grateful to R. A. Konoplya
for pointing out in Ref. \cite{Konoplya} the typo in the numerical
factor of the expression of $a^{r}$ in the first version of our
manuscript. }
\begin{eqnarray}
 \phi &=&\frac{1}{2}\ln f(r), \label{phi_spherically}\\
 a^\mu &=&\left(0,-\frac{1}{2}f'(r),0,0\right), \label{a_spherically}\\
 T &=& \frac{{\hbar}}{4\pi}|f'(r)|. \label{T_spherically}
\end{eqnarray}
The energy on the screen is
\begin{eqnarray}
 E = \frac{1}{2}|f'(r)|r^2. \label{E_spherically}
\end{eqnarray}

Next, we focus on the Schwarzschild black hole and the RN black
hole.

\subsection{The Schwarzschild black hole}

For the Schwarzschild black hole, the metric has the form of
(\ref{general1}) with
\begin{equation}
 f(r)=1-\frac{2M}{r}.\label{f(r)_Schwarz}
\end{equation}
The event horizon of the balk hole is located at $r_{EH}=2M$, and
the Hawking temperature $T_H$ on the event horizon is given by
\begin{eqnarray}
T_H=\frac{{\hbar}}{4\pi r_{EH}}. \label{Schwarzschild_TH}
\end{eqnarray}

The acceleration and the Unruh-Verlinde temperature on the screen are
calculated as
\begin{eqnarray}
 a^\mu &=&\left(0,\frac{M}{r^2},0,0\right), \label{a_Schwarzschild}\\
 T &=& \frac{{\hbar}M}{2\pi r^2}. \label{T_Schwarzschild}
\end{eqnarray}
Note that the Unruh-Verlinde temperature on the event horizon
$r=r_{EH}$ is just the Hawking temperature $T_H$:
\begin{eqnarray}
 T|_{r=r_{EH}}=\frac{\hbar}{8\pi M}=T_H. \label{RelationWithTh_Sch}
\end{eqnarray}
Then, with the expressions (\ref{E_spherically}) and (\ref{f(r)_Schwarz}), we
obtain the energy on the screen:
\begin{eqnarray}
 E = M. \label{E_Schwarzschild}
\end{eqnarray}
It is interesting to note that this energy is independent of the
radius of the screen. In fact this is a consequence of the Gauss'
law.

\subsection{The RN black hole}

As an important example of spherically symmetric black holes, the
RN black hole has the metric form of (\ref{general1}) with
\begin{equation}
f(r)=\left(1-\frac{2M}{r}+\frac{Q^2}{r^2}\right). \label{f(r)_RN}
\end{equation}
There are two horizons, the event horizon with radius $r_+$ and the Cauchy
horizon with radius $r_{-}$, where
\begin{equation}\label{horizonRN}
r_{\pm}= M \pm \sqrt{M^2 -Q^2}.
\end{equation}
The extremal RN black hole corresponds to the case $r_{+} = r_{-}$ or,
equivalently, $M=Q$. The Hawking temperatures $T_{H\pm}$ on the horizons are
\begin{eqnarray}
T_{H\pm}=\frac{{\hbar}}{4\pi}\frac{(r_+ -r_-)}{r_{\pm}^2}
   =\frac{{\hbar}}{2\pi}\frac{\sqrt{M^2 -Q^2}}{r_{\pm}^2}.
\end{eqnarray}

The potential, the acceleration and the Unruh-Verlinde temperature
on the screen read
\begin{eqnarray}
 \phi &=&\frac{1}{2}\ln\left(1-\frac{2M}{r}+\frac{Q^2}{r^2}\right),
         \label{phi_RN}\\
 a^\mu &=&\left(0,\frac{Q^2-Mr}{r^3},0,0\right). \label{a_RN} \\
 T &=&  \frac{{\hbar}}{2\pi}
      \frac{|Mr-Q^2|}{r^3}. \label{T_RN}
\end{eqnarray}
On the horizons $r=r_{\pm}$, we have
\begin{eqnarray}
 T|_{r=r_{\pm}}=\frac{{\hbar}}{2\pi }
      \frac{\sqrt{M^2 -Q^2}}{r_{\pm}^2},
\end{eqnarray}
which shows that the Unruh-Verlinde temperatures on both the horizons equal to
the Hawking temperatures $T_{H\pm}$. The solution of $\partial_r T=0$ is
$r=3q^2/(2M)\equiv r_0$. So for $r_0<r_+$, i.e., $Q^2<8M^2/9$, the maximum of
$T$ in the range $r\geq r_+$ is at $r=r_+$. For $8M^2/9<Q^2<M^2$, the maximum
of $T$ in the range $r\geq r_+$ is at $r=r_0(>r_+)$ and is given by
\begin{eqnarray}
 T_{max}=T|_{r_0}=\frac{{\hbar}}{2\pi }
      \frac{2M^3}{27Q^4}.
\end{eqnarray}

The energy on the screen is
\begin{eqnarray}
 E = \left|M-\frac{Q^2}{r}\right|, \label{E_RN}
\end{eqnarray}
which is the well-known Komar energy of the RN black hole. Now this energy is
dependent on the radius of the screen, which is also a consequence of the
Gauss' laws of gravity and electrostatics. For the screen at infinity, the
energy is reduced to that of the Schwarzschild black hole: $E=M$. It is
interesting to note that the energies on the horizons $r_{\pm}$ are the same:
\begin{eqnarray}
 E|_{r=r_{\pm}} = \sqrt{M^2-Q^2}. \label{E_RN_h}
\end{eqnarray}

For the extremal RN black hole with $r_+=r_-$, the corresponding
Unruh-Verlinde temperature and energy are
\begin{eqnarray}
 T &=& \frac{{\hbar}}{2\pi}
      \frac{M|r-M|}{r^3},\\
 E &=& M\left|1-\frac{M}{r}\right|. \label{E_RN2}
\end{eqnarray}
They vanish at the horizon $r_+=M$. This result is in agreement with the
statements that the extremal black hole has a unique internal state
\cite{Hawking1995} and its temperature is zero due to the vanishing of surface
gravity on horizon.


\section{Unruh-Verlinde Temperature and energy for the Kerr black hole}
 \label{Sec4}

In this section, we extend this work to the Kerr black hole. The Kerr solution
describes both the stationary axisymmetric asymptotically flat gravitational
field outside a massive rotating body and a rotating black hole with mass and
angular momentum. The Kerr black hole can also be viewed as the final state of
a collapsing star, uniquely determined by its mass and rate of rotation.
Moreover, its thermodynamical behavior is very different from the Schwarzschild
black hole and the RN black hole, because of its much more complicated causal
structure. Hence its study is of great interest in understanding physical
properties of astrophysical objects, as well as in checking any conjecture
about thermodynamical properties of black holes.

In terms of Boyer--Lindquist coordinates, the Euclidean Kerr metric
reads
\begin{eqnarray}
 ds^{2} &=& -\left(1-\frac{2Mr}{\varrho^{2}}\right)dt^2
           + \frac{\varrho^{2}}{\Delta}dr^{2}
           +\varrho^{2}d\theta^{2} \nonumber\\
      && +\left((a^2+r^2)\sin^{2}\theta
                +\frac{2Mra^2\sin^{4}\theta}{\varrho^{2}}
          \right)d\varphi^2    -\frac{4Mra\sin^{2}\theta}{\varrho^{2}}dtd\varphi,
          \label{kermet}
\end{eqnarray}
where $\Delta$ is the Kerr horizon function
\begin{eqnarray}
\Delta = r^{2}-2Mr +a^{2},
\end{eqnarray}
and
\begin{eqnarray}
\varrho = r^{2} + a^{2}\cos^{2}\theta.
\end{eqnarray}
Here $a$ is the angular momentum for unit mass as measured from
the infinity; it vanishes in the Schwarzschild limit. The
nonextremal Kerr black hole has the event horizon $r_{+}$ and the
Cauchy horizon $r_{-}$ at
\begin{equation}
r_{\pm}=M \pm \sqrt{M^2-a^2}.
\end{equation}
The extreme case corresponds to $r_{+}=r_{-}$ or $M=a$. The Hawking
temperatures on the horizons are
\begin{eqnarray}
T_{H\pm}=\frac{{\hbar}}{4\pi}\frac{(r_+ -r_-)}{(r_{\pm}^2+a^2)}
   =\frac{{\hbar}}{4\pi}\frac{\sqrt{M^2 -a^2}}{Mr_{\pm}}.
\end{eqnarray}

\begin{figure}[h]
\begin{center}
\subfigure[$r\geq r_+=16$]{\label{fig:ContourPhiAOut}
\includegraphics[width=6.5cm]{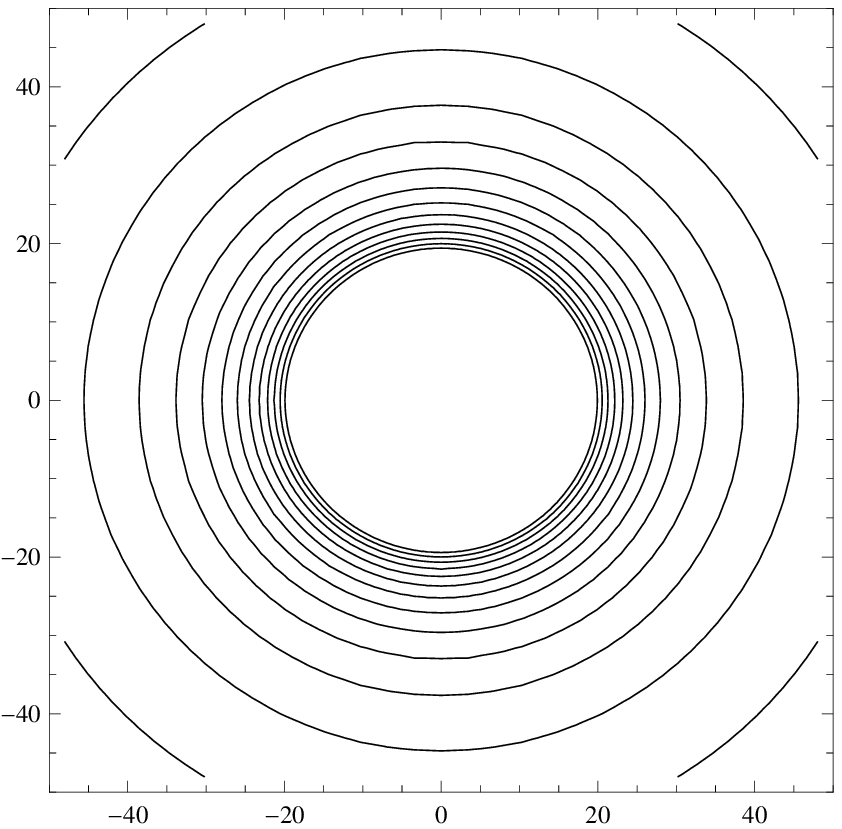}}
\subfigure[$r\leq r_-=4$]{\label{fig:ContourPhiAIn}
\includegraphics[width=6.5cm]{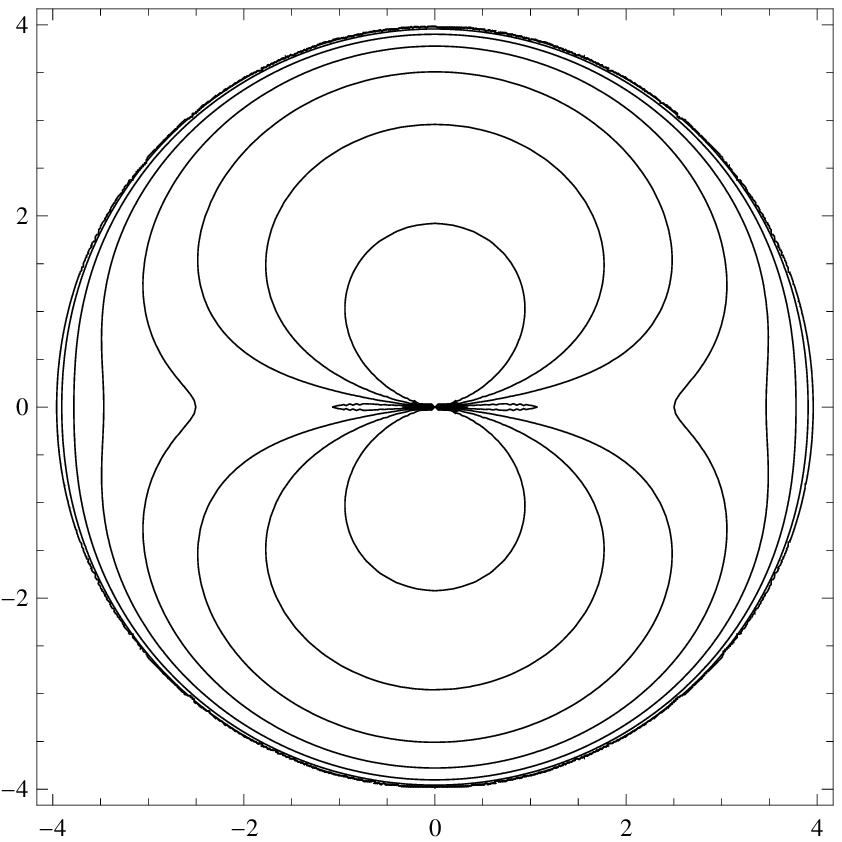}}
\end{center}\vskip -4mm
\caption{  A contour plot of the potential $\phi$
 in the $y$-$z$ plane for the Kerr
black hole. The parameters are set to $M=10$ and $a=8$.}
\label{fig:ContourPhiA}
\end{figure}

\begin{figure}[h]
\begin{center}
\subfigure[$r\geq r_+=11.4$]{\label{fig:ContourPhiBOut}
\includegraphics[width=6.5cm]{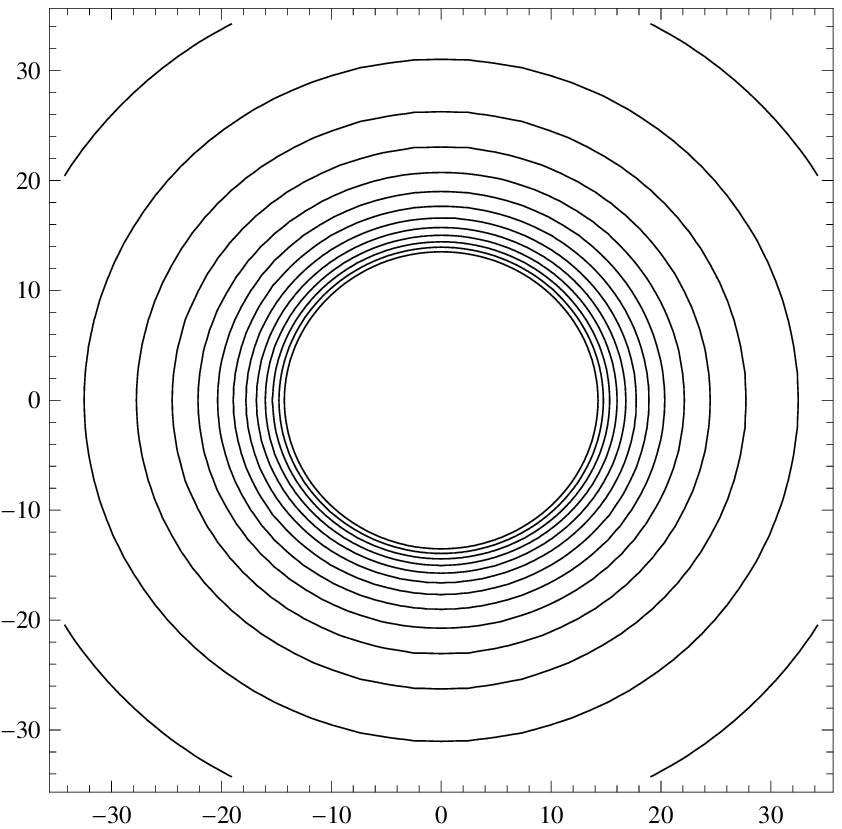}}
\subfigure[$r\leq r_-=8.6$]{\label{fig:ContourPhiAIn}
\includegraphics[width=6.5cm]{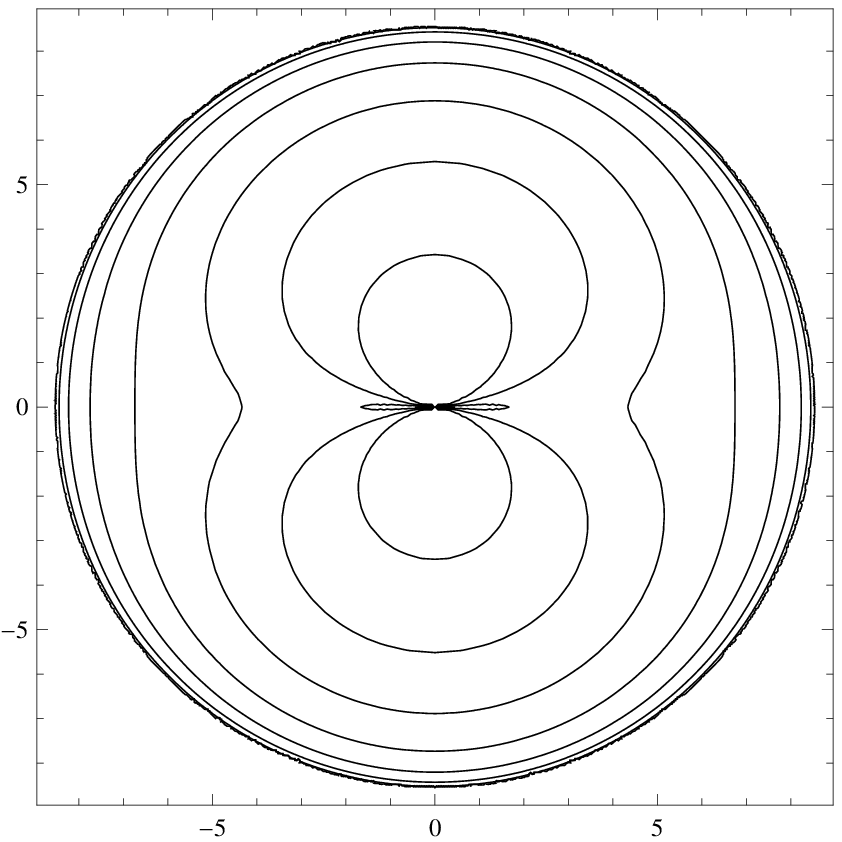}}
\end{center}\vskip -4mm
\caption{  A contour plot of the potential $\phi$
  in the $y$-$z$ plane for the Kerr
black hole with $a\rightarrow M$. The parameters are set to
$M=10$ and $a=9.9$.}\label{fig:ContourPhiB}
\end{figure}

The solution for the timelike Killing vector $\xi_\mu$ is read as
\cite{WuXN1002.1275}
\begin{eqnarray}
  \xi_\mu=\left(-1+\frac{2Mr(\Omega-2a^2 Mr\sin^2\theta)}
                        {\Omega\varrho^{2}},0,0,0\right),
 \label{KillingVectorKerr}
\end{eqnarray}
where $\Omega=(r^2+a^2)^2-a^2{\Delta}\sin^2\theta$.  The corresponding
potential is
\begin{eqnarray}
  \phi=-\frac{1}{2}\log\left({1+\frac{2Mr(r^2+a^2)}{\Delta\varrho^2}}\right),
 \label{PotentialKerr}
\end{eqnarray}
which shows that equipotential surfaces are dependent with two parameters $r$
and $\theta$, hence the holographic screens are not spherically symmetric but
axisymmetric. However, when $r\gg r_+$ or $r\rightarrow r_{\pm}$, the
holographic screens are approximate spherically symmetric, which can be seen
from the contour plots of the potential $\phi$ in the $y$-$z$ plane showed in
Figs. \ref{fig:ContourPhiA} and \ref{fig:ContourPhiB}. It can be seen that,
when $a\rightarrow M$, the effect of the angular momentum are remarkable near
the horizons. When far away from the event horizon, the equipotential line can
be approximated as a spherical surface.


\begin{figure}[h]
\begin{center}
\subfigure[$r\geq r_+=16$]{\label{fig:ContourTAout}
\includegraphics[width=6.5cm]{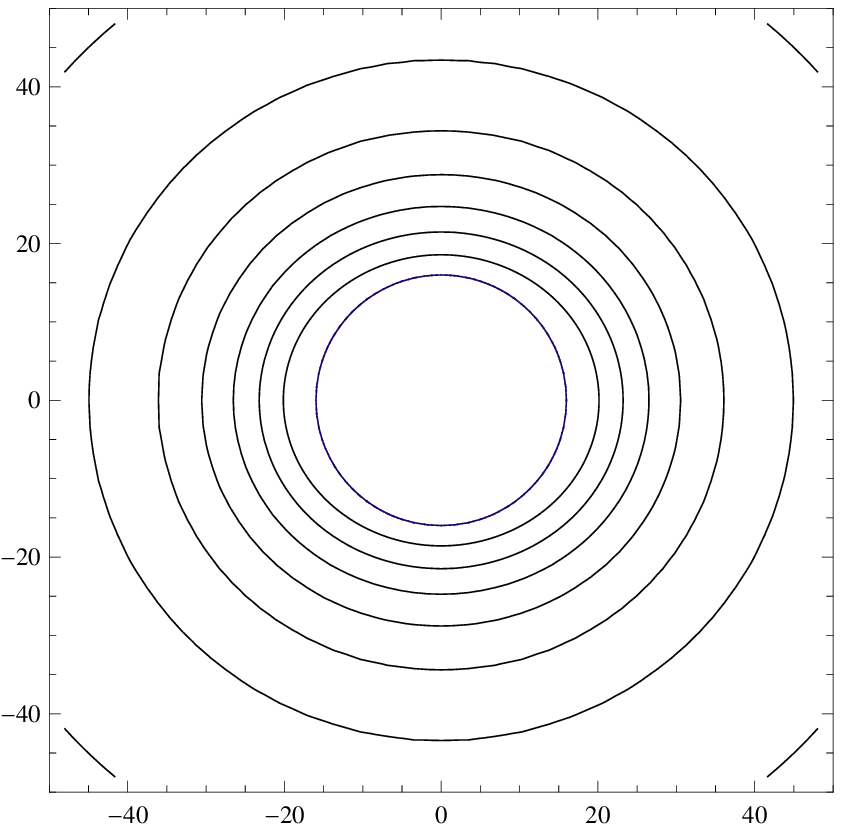}}
\subfigure[$r\leq r_-=4$]{\label{fig:ContourTAin}
\includegraphics[width=6.5cm]{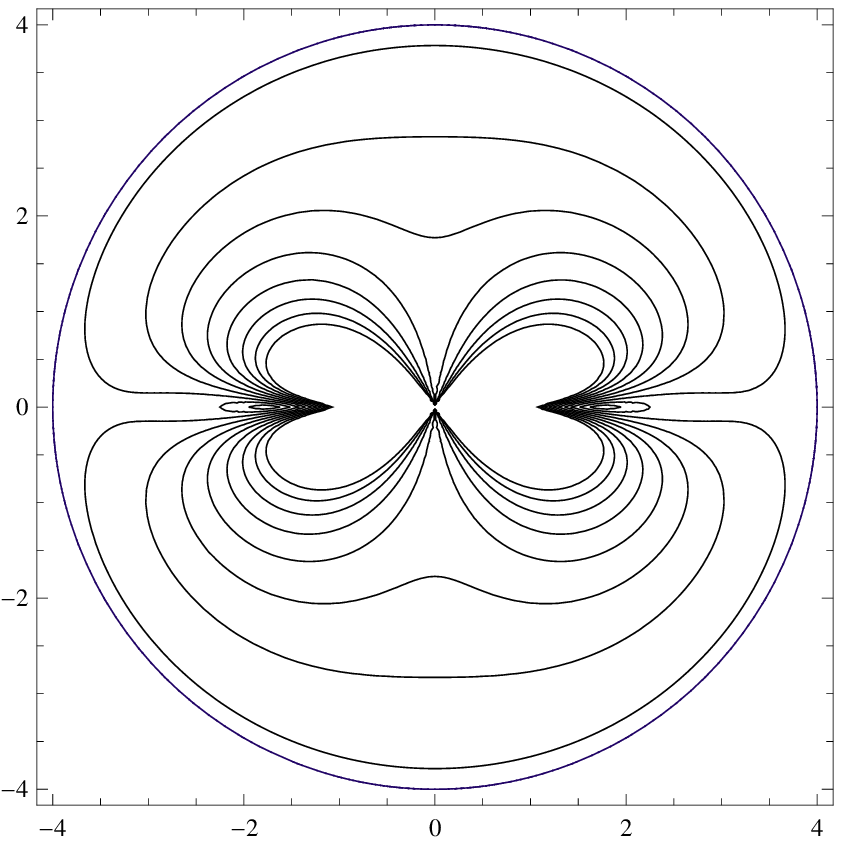}}
\end{center}\vskip -4mm
\caption{  A contour plot of the temperature $T$
  in the $y$-$z$ plane for the Kerr black
hole. The parameters are set to $M=10$ and
$a=8$.}\label{fig:ContourTA}
\end{figure}

\begin{figure}[h]
\begin{center}
\subfigure[$r\geq r_+=11.4$]{\label{fig:ContourTBout}
\includegraphics[width=6.5cm]{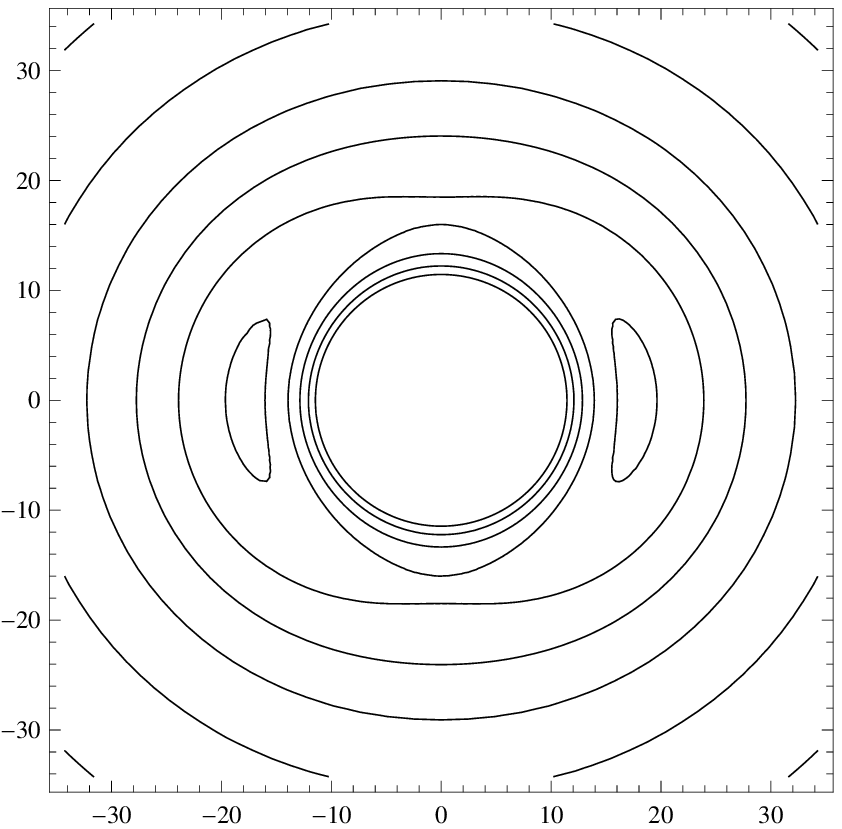}}
\subfigure[$r\leq r_-=8.6$]{\label{fig:ContourTBin}
\includegraphics[width=6.5cm]{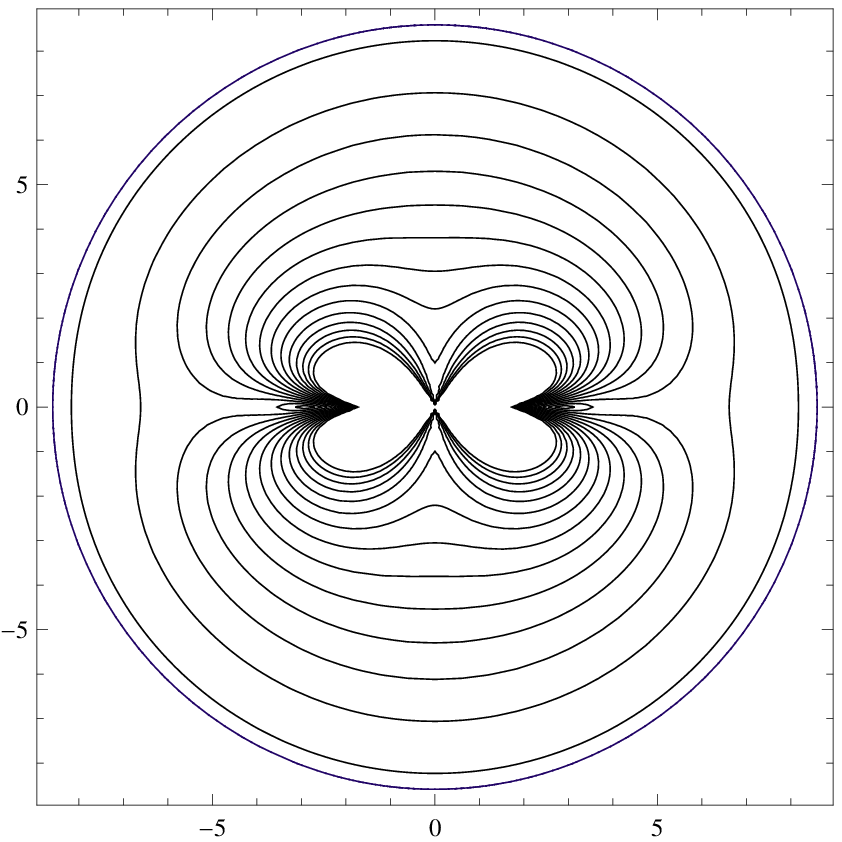}}
\end{center}\vskip -4mm
\caption{  A contour plot of the temperature $T$
  in the $y$-$z$ plane for the Kerr black hole
with $a\rightarrow M$. The parameters are set to $M=10$ and
$a=9.9$.}\label{fig:ContourTB}
\end{figure}

The non-zero components of the acceleration are
\begin{eqnarray}
  a^r &=& \frac{2Mr(a^2+r^2)
            \left(M-r-\frac{\Delta r}{\varrho^2}\right)
            +M\Delta(a^2+3r^2)}
         {\varrho^2[\Delta\varrho^2+2Mr(a^2+r^2)]}, \nonumber \\
  a^\theta &=& \frac{Mra^2(a^2+r^2)\sin(2\theta)}
           {\varrho^4[2Mr(a^2+r^2)+\Delta\varrho^2]}.
           \label{a_Kerr}
\end{eqnarray}
The Unruh-Verlinde temperature is given by
\begin{eqnarray}
T&=&\frac{h}{2\pi } \left\{
  \frac{M^2\big[\left(a^2+3 r^2\right) \Delta  \varrho ^2+2 r
              \left(a^2+r^2\right)
              \left(M \varrho ^2-r \left(\Delta
                    +\varrho^2\right)\right)\big]^2 }
       {\varrho ^4 \left[2 M r \left(a^2+r^2\right)
                +\Delta \varrho ^2\right]^3} \right. \nonumber \\
  && \quad\quad~ + \left. \frac{\Delta
          \left[a^2 M r (a^2+r^2)
               \texttt{sin}(2\theta)
          \right]^2 }
       {\varrho ^4 \left[2 M r \left(a^2+r^2\right)
                +\Delta \varrho ^2\right]^3} \right\}
      ^{\frac{1}{2}}          .
         \label{T_Kerr}
\end{eqnarray}
On the horizons $r=r_{\pm}$ ($\Delta=0$), the Unruh-Verlinde temperatures
become
\begin{eqnarray}
T|_{r=r_{\pm}}&=&\left. \frac{\hbar}{2\pi }\sqrt{
  \frac{(M -r)^2 }{2 M r (a^2+r^2)} }~ \right|_{r=r_{\pm}}\nonumber \\
  &=& \frac{{\hbar}}{4\pi}
    \frac{\sqrt{M^2-a^2}} {Mr_{\pm}}
    =T_{H\pm}.
         \label{T_Kerr_horizons}
\end{eqnarray}
So, the Unruh-Verlinde temperatures on both horizons are equal to the Hawking
temperatures. For vanishing $a$, the above result is reduced to
$T|_{r=r_+}=\frac{{\hbar}}{4\pi r_+}$, which is just the Hawking temperature
(\ref{T_Schwarzschild}) on the horizon of the Schwarzschild black hole. Contour
plots of the temperature $T$ in the $y$-$z$ plane are shown in Figs.
\ref{fig:ContourTA} and \ref{fig:ContourTB}.

Now, by using the equipartition law of energy and the holographic principle
$E={1\over 2} \int_{\cal S}  T dN=\frac{1}{2\hbar} \int_{\cal S}  T dA$,
we get the energies on the horizons
\begin{eqnarray}
 E|_{r=r_{\pm}} 
 &=& \frac{\sqrt{M^2-a^2}}{2M r_{\pm}}\left( r_{\pm}^2+a^2 \right) \nonumber \\
 &=& \sqrt{M^2-a^2}, \label{E_Kerr_h}
\end{eqnarray}
which is the reduced mass $M_0$ of the Kerr black hole. When far away from the
event horizon, or equivalently $r\gg r_+$, the screen is approximated as a
sphere and the energy can be expressed as
\begin{eqnarray}
 E
   &\approx&  M\left(1-4\frac{a^2M}{r^3}\right). \label{E_Kerr_Approx}
\end{eqnarray}
For $r\rightarrow\infty$, the result is $E=M$, which is independent of the
angular of the Kerr black hole. For $a=0$, the expression (\ref{E_Kerr_Approx})
reduce to $E=M$, which is just the energy expression (\ref{E_Schwarzschild})
for the Schwarzschild black hole.

\section{Discussions}\label{SecConclusions}

In this paper, with the holographic principle and the equipartition law of
energy, we investigate the Unruh-Verlinde temperature and energy on holographic
screens for several 4-dimensional black holes with static spherically symmetric
metric and stationary axisymmetric metric.

On the event horizon of a static spherically symmetric black hole, the Hawking
temperature $T_H$ is related to the surface gravity
$\kappa=\frac{1}{2}f'(r_{EH})$ by the relation
$T_H=\frac{\hbar\kappa}{2\pi}=\frac{{\hbar}}{4\pi}f'(r_{EH})$, while the
Unruh-Verlinde temperatures $T$ is related to the acceleration $a^\mu
=\left(0,-\frac{f'(r)}{2},0,0\right)$ by $T=-{\hbar \over 2\pi} e^{\phi} n^\mu
a_\mu=\frac{{\hbar}}{4\pi}f'(r_{EH})$, which shows that the Unruh-Verlinde
temperatures $T$ for a static spherically symmetric black hole is identical to
the Hawking temperature. Hence, outside the horizons, the Unruh-Verlinde
temperature $T$ can be considered as a generalized Hawking temperature on the
holographic screen. For the Kerr black case, we also consider $T$ as the
generalized Hawking temperature. On the screen located at the infinity, the
surface gravity or the acceleration of a particle (which is free there)
vanishes, hence the temperature is zero.

The entropic force (\ref{Fmu}) for a static spherically symmetric black hole
can be calculated as $F_{\mu}=(0,-\frac{mf'(r)}{2\sqrt{f(r)}},0,0)$. The
magnitude of the force is
$F=\sqrt{g^{\mu\nu}F_{\mu}F_{\nu}}=\frac{1}{2}mf'(r)$. For the Schwarzschild
black hole, the entropic force $F$ is just the Newton force
\begin{equation}
 F=\frac{GmM}{r^2}, \label{F_Schwarzschild}
\end{equation}
here we recover the Newton's gravitational constant $G$. While, for the RN
case, the entropic force is turned out to be
\begin{equation}
 F=\frac{GmM}{r^2}\left(1-\frac{Q^2}{Mr}\right), \label{F_RN}
\end{equation}
which shows that there is a corrected term caused by the charge of the black
hole. This force is related to the pure gravitational effect. The gravity or
the geometry near the event horizon of the black hole is affected by the energy
of the electric field. When far away form the horizon, where the electric field
becomes weak, the entropic force will recover to the Newton one. Note that,
since $r\geq r_+>M>Q$, we have $Q^2/(Mr)<1$ and $F>0$.  Similarly, the angular
momentum of the Kerr black hole will also affect the structure of the spacetime
near the event horizon. The entropic force of the Kerr black hole is read as
\begin{eqnarray}
 F=\frac{Gm\sqrt{M^2-a^2}}{r^2}\label{F_Kerr_1}
\end{eqnarray}
and
\begin{eqnarray}
 F\approx\frac{GmM}{r^2}\left(1-4\frac{a^2M}{r^3}\right)
  \label{F_Kerr_2}
\end{eqnarray}
at $r=r_+$ and $r\gg r_+$, respectively.  It also reduces at large $r$ to the
Newton force because the effect of the ``field" caused by the angular momentum
of the black hole becomes weaker with the increase of $r$.

The energy got from the holographic principle is indeed the Komar energy on the
screen. With the same explanation above, we also can understand that the energy
$E$ on the screen will trend to the ADM mass $M$ when $r\rightarrow\infty$ for
anyone of the black holes considered in the paper. This is because the ``field"
produced by the charge or the angular momentum is strong on the screen near the
event horizon and weak when far away from the black hole. On the screen at
infinity, the effect of them vanishes and the energy is naturally equal to the
black hole mass $M$ according to the gravitational Gauss' law. We see clearly
from (\ref{E_RN_h}) and (\ref{E_Kerr_h}) that when applying the entropic force
idea to the black hole horizon, it is the reduced mass $M_0$ that takes the
place of the black hole mass $M$. For the Schwarzschild black hole, there is no
other parameter beside $M$, so the reduced mass $M_0$ is just the black hole
mass $M$. For the RN black hole, the reduced mass is read from (\ref{E_RN}) as
$M_0=M-\frac{Q^2}{r}$, which is $M_0=\sqrt{M^2-Q^2}$ on the event horizon. For
the Kerr case, the reduced mass is $M_0=\sqrt{M^2-a^2}$ on the event horizon
and is $M_0 {\approx} M \left(1-4\frac{a^2M}{r^3}\right)$ at $r\gg r_+$.
Therefore, with the black hole mass $M$ replaced by the reduced mass $M_0$, we
can rewrite the entropic force of a black hole as
\begin{equation}
 F=\frac{GmM_0}{r^2}, \label{F_Entropic}
\end{equation}
which has the same form as the Newton's law of gravity. This formula shows that
the force between two neutral particles is exactly the Newton's force of
gravity, while the force between a neutral particle and a charged particle
would departure from the Newton's law of gravity. Hence, the idea of entropic
force could be tested by experiments according to Eq. (\ref{F_Entropic}).
Explicitly, the force between a neutral particle and a charged particle is $F=\frac{Gm}{r^2}\left(M-\frac{Q^2}{r}\right)$ according entropic force idea. Therefore, by comparing the high accurate experiment results of the force between a neutral particle and a charged particle with the theoretical predictions of the entropic force, one could check whether the entropic idea is right or wrong.


\section*{Acknowledgment}


This work was supported by the Program for New Century Excellent Talents in
University, the National Natural Science Foundation of China (No. 10705013),
the Key Project of Chinese Ministry
of Education (No. 109153), and the Fundamental Research Funds for the Central
Universities (No. lzujbky-2009-54, No. lzujbky-2009-122 and No.
lzujbky-2009-163).


%

\end{document}